\pdfoutput=1
\documentclass[12pt,a4paper]{article}

\usepackage{ifthen}
\newboolean{pdflatex}
\setboolean{pdflatex}{true}

\newboolean{articletitles}
\setboolean{articletitles}{true}

\newboolean{uprightparticles}
\setboolean{uprightparticles}{false}

\def\paperauthors{LHCb collaboration}
\def\paperasciititle{Associated Z + J/psi production as a probe of multiparton interactions in the forward region}
\def\papertitle{Associated $Z$ + $J/\psi$ production as a probe of multiparton interactions in the forward region}
\def\paperkeywords{{High Energy Physics}, {LHCb}}
\def\papercopyright{\the\year\ CERN for the benefit of the LHCb collaboration}
\def\paperlicence{CC BY 4.0 licence}
\def\paperlicenceurl{https://creativecommons.org/licenses/by/4.0/}

\newif\ifEnableSectionTOCLinks
\EnableSectionTOCLinksfalse

\usepackage[top=1in, bottom=1.25in, left=1in, right=1in]{geometry}

\usepackage{microtype}
\usepackage{xspace}
\usepackage{caption}

\usepackage{graphicx}
\usepackage{color}
\usepackage{colortbl}
\graphicspath{{./figs/}}

\usepackage{amsmath}
\usepackage{amssymb}
\usepackage{amsfonts}
\usepackage{upgreek}

\usepackage[pdftex,
            pdfauthor={\paperauthors},
            pdftitle={\paperasciititle},
            pdfkeywords={\paperkeywords}]{hyperref}
\usepackage{hyperxmp}
\hypersetup{
    pdfcopyright={Copyright (C) \papercopyright},
    pdflicenseurl={\paperlicenceurl}
}

\usepackage[bottom,flushmargin,hang,multiple]{footmisc}

\usepackage[all]{hypcap}

\usepackage{xspace} 
\usepackage{upgreek}

\def\lhcb   {\mbox{LHCb}\xspace}

\def\MagUp {\mbox{\em Mag\kern -0.05em Up}\xspace}

\ifthenelse{\boolean{uprightparticles}}%
{

 \def\Pmu         {\ensuremath{\upmu}\xspace}

 \def\Ppsi        {\ensuremath{\uppsi}\xspace}

 \def\PDelta      {\ensuremath{\Delta}\xspace}                 
 \def\PXi         {\ensuremath{\Xi}\xspace}                 
 \def\PLambda     {\ensuremath{\Lambda}\xspace}                 
 \def\PSigma      {\ensuremath{\Sigma}\xspace}                 
 \def\POmega      {\ensuremath{\Omega}\xspace}                 
 \def\PUpsilon    {\ensuremath{\Upsilon}\xspace}
 \let\oldPi\Pi
 \def\PPi         {\ensuremath{\oldPi}\xspace}

 \def\PB      {\ensuremath{\mathrm{B}}\xspace}                 
 \def\PD      {\ensuremath{\mathrm{D}}\xspace}                 
 \def\PJ      {\ensuremath{\mathrm{J}}\xspace}                 
 \def\PK      {\ensuremath{\mathrm{K}}\xspace}                 
 \def\PZ      {\ensuremath{\mathrm{Z}}\xspace}                 
 \def\Pb      {\ensuremath{\mathrm{b}}\xspace}                 
 \def\Pc      {\ensuremath{\mathrm{c}}\xspace}

 \def\Pp      {\ensuremath{\mathrm{p}}\xspace}                 

 \def\Ps      {\ensuremath{\mathrm{s}}\xspace}

 \def\thebaroffset{0.0em}
}
{

 \def\Pmu         {\ensuremath{\mu}\xspace}

 \def\Ppsi        {\ensuremath{\psi}\xspace}                 
                  
 \mathchardef\PDelta="7101
 \mathchardef\PXi="7104
 \mathchardef\PLambda="7103
 \mathchardef\PSigma="7106
 \mathchardef\POmega="710A
 \mathchardef\PUpsilon="7107
 \mathchardef\PPi="7105
 \def\PB      {\ensuremath{B}\xspace}                 
 \def\PD      {\ensuremath{D}\xspace}                 
 \def\PJ      {\ensuremath{J}\xspace}                 
 \def\PK      {\ensuremath{K}\xspace}                 
 \def\PZ      {\ensuremath{Z}\xspace}                 
 \def\Pb      {\ensuremath{b}\xspace}                 
 \def\Pc      {\ensuremath{c}\xspace}

 \def\Pp      {\ensuremath{p}\xspace}                 

 \def\Ps      {\ensuremath{s}\xspace}

 \def\thebaroffset{0.18em}
}
\newcommand{\offsetoverline}[2][\thebaroffset]{\kern #1\overline{\kern -#1 #2}}%

\makeatletter
\ifcase \@ptsize \relax
  \newcommand{\miniscule}{\@setfontsize\miniscule{4}{5}}
\or
  \newcommand{\miniscule}{\@setfontsize\miniscule{5}{6}}
\or
  \newcommand{\miniscule}{\@setfontsize\miniscule{5}{6}}
\fi
\makeatother

\DeclareRobustCommand{\optbar}[1]{\shortstack{{\miniscule (\rule[.5ex]{1.25em}{.18mm})}
  \\ [-.7ex] $#1$}}

\def\mumu       {{\ensuremath{\Pmu^+\Pmu^-}}\xspace}

\def\Z      {{\ensuremath{\PZ}}\xspace}

\def\squark    {{\ensuremath{\Ps}}\xspace}

\def\cquark    {{\ensuremath{\Pc}}\xspace}

\def\bquark    {{\ensuremath{\Pb}}\xspace}

\def\KorKbar {\kern \thebaroffset\optbar{\kern -\thebaroffset \PK}{}\xspace}

\def\D       {{\ensuremath{\PD}}\xspace}

\def\DorDbar {\kern \thebaroffset\optbar{\kern -\thebaroffset \PD}\xspace}

\def\Dp      {{\ensuremath{\D^+}}\xspace}
\def\Dm      {{\ensuremath{\D^-}}\xspace}

\def\DpDm    {\ensuremath{\Dp {\kern -0.16em \Dm}}\xspace}

\def\B       {{\ensuremath{\PB}}\xspace}

\def\BorBbar {\kern \thebaroffset\optbar{\kern -\thebaroffset \PB}\xspace}

\def\Bd      {{\ensuremath{\B^0}}\xspace}

\def\BdorBdbar {\kern \thebaroffset\optbar{\kern -\thebaroffset \Bd}\xspace}

\def\Bs      {{\ensuremath{\B^0_\squark}}\xspace}

\def\BsorBsbar {\kern \thebaroffset\optbar{\kern -\thebaroffset \Bs}\xspace}

\def\jpsi     {{\ensuremath{{\PJ\mskip -3mu/\mskip -2mu\Ppsi}}}\xspace}

\def\Upsilonres  {{\ensuremath{\PUpsilon}}\xspace}
\def\Y#1S{\ensuremath{\PUpsilon{(#1S)}}\xspace}

\def\proton      {{\ensuremath{\Pp}}\xspace}
\def\antiproton  {{\ensuremath{\overline \proton}}\xspace}

\def\LorLbar     {\kern \thebaroffset\optbar{\kern -\thebaroffset \PLambda}\xspace}

\def\BF         {{\ensuremath{\mathcal{B}}}\xspace}
\def\BR         {\BF}

\newcommand{\decay}[2]{\ensuremath{\mathinner{#1\!\to #2}}\xspace}

\def\to                 {\ensuremath{\rightarrow}\xspace}

\def\AT#1     {\ensuremath{A_{\mathrm{T}}^{#1}}\xspace}           

\def\C#1      {\ensuremath{\mathcal{C}_{#1}}\xspace}                       
\def\Cp#1     {\ensuremath{\mathcal{C}_{#1}^{'}}\xspace}                    
\def\Ceff#1   {\ensuremath{\mathcal{C}_{#1}^{\mathrm{(eff)}}}\xspace}        
\def\Cpeff#1  {\ensuremath{\mathcal{C}_{#1}^{'\mathrm{(eff)}}}\xspace}       
\def\Ope#1    {\ensuremath{\mathcal{O}_{#1}}\xspace}                       
\def\Opep#1   {\ensuremath{\mathcal{O}_{#1}^{'}}\xspace}                    

\newcommand{\aunit}[1]{\ensuremath{\text{\,#1}}}       

\newcommand{\tev}{\aunit{Te\kern -0.1em V}\xspace}
\newcommand{\gev}{\aunit{Ge\kern -0.1em V}\xspace}
\newcommand{\mev}{\aunit{Me\kern -0.1em V}\xspace}
\newcommand{\kev}{\aunit{ke\kern -0.1em V}\xspace}
\newcommand{\ev}{\aunit{e\kern -0.1em V}\xspace}
 
\newcommand{\mevc}{\ensuremath{\aunit{Me\kern -0.1em V\!/}c}\xspace}
\newcommand{\gevc}{\ensuremath{\aunit{Ge\kern -0.1em V\!/}c}\xspace}
\newcommand{\mevcc}{\ensuremath{\aunit{Me\kern -0.1em V\!/}c^2}\xspace}
\newcommand{\gevcc}{\ensuremath{\aunit{Ge\kern -0.1em V\!/}c^2}\xspace}

\def\mbarn{\aunit{mb}\xspace}

\def\pb {\aunit{pb}\xspace}

\def\fb   {\ensuremath{\aunit{fb}}\xspace}
\def\invfb   {\ensuremath{\fb^{-1}}\xspace}

\newcommand{\chisqndf}{\ensuremath{\chi^2/\mathrm{ndf}}\xspace}

\def\deriv {\ensuremath{\mathrm{d}}}

\def\gsim{{~\raise.15em\hbox{$>$}\kern-.85em
          \lower.35em\hbox{$\sim$}~}\xspace}
\def\lsim{{~\raise.15em\hbox{$<$}\kern-.85em
          \lower.35em\hbox{$\sim$}~}\xspace}

\def\sPlot{\mbox{\em sPlot}\xspace}

\def\sqs   {\ensuremath{\protect\sqrt{s}}\xspace}

\def\pt         {\ensuremath{p_{\mathrm{T}}}\xspace}

\def\ptot       {\ensuremath{p}\xspace}

\def\evtgen     {\mbox{\textsc{EvtGen}}\xspace}

\def\geant      {\mbox{\textsc{Geant4}}\xspace}

\def\photos     {\mbox{\textsc{Photos}}\xspace}

\def\pythia     {\mbox{\textsc{Pythia}}\xspace}

\def\tell1  {TELL1\xspace}
\def\ukl1   {UKL1\xspace}

\newcommand{\ie}{\mbox{\itshape i.e.}\xspace}

\newcommand{\lhcborcid}[1]{\href{https://orcid.org/#1}{\hspace*{0.1em}\raisebox{-0.45ex}{\includegraphics[width=1em]{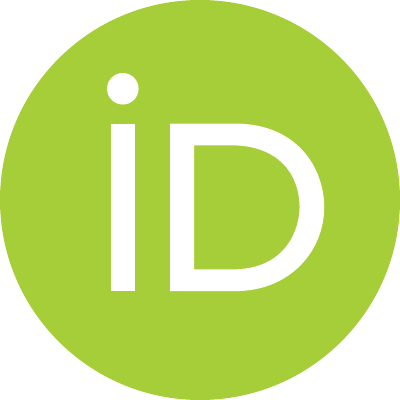}}}}

\hypersetup{
  colorlinks   = true,
  urlcolor     = blue,
  linkcolor    = blue,
  citecolor    = red
}

\ifEnableSectionTOCLinks
    \usepackage[explicit]{titlesec}
    
    \let\oldcontentsline\contentsline
    \renewcommand

    \titleformat{\section}{\normalfont\Large\bf}{\hyperlink{tocsection.\thesection}{{\thesection} \parbox[t]{\dimexpr\textwidth-1pc}{#1}}}{1pc}{}

    \titleformat{\subsection}{\normalfont\bf}{\hyperlink{tocsubsection.\thesubsection}{{\thesubsection} \parbox[t]{\dimexpr\textwidth-1pc}{#1}}}{1pc}{}

    \titleformat{name=\section,numberless}[display]{}{}{0pt}{\normalfont\Huge\bfseries #1}
\fi

\usepackage{cite}
\usepackage{LHCb/mciteplus}

\def\zpt{\ensuremath{p^{\Z}_{\mathrm{T}}}\xspace}
\def\jpsipt{\ensuremath{p^{\jpsi}_{\mathrm{T}}}\xspace}

\def\mupt{\ensuremath{p^{\mu}_{\mathrm{T}}}\xspace}

\begin{document}

\renewcommand{\thefootnote}{\fnsymbol{footnote}}
\setcounter{footnote}{1}

\begin{titlepage}
\pagenumbering{roman}

\vspace*{-1.5cm}
\centerline{\large EUROPEAN ORGANIZATION FOR NUCLEAR RESEARCH (CERN)}
\vspace*{1.5cm}
\noindent
\begin{tabular*}{\linewidth}{lc@{\extracolsep{\fill}}r@{\extracolsep{0pt}}}
\ifthenelse{\boolean{pdflatex}}
{\vspace*{-1.5cm}\mbox{\!\!\!\includegraphics[width=.14\textwidth]{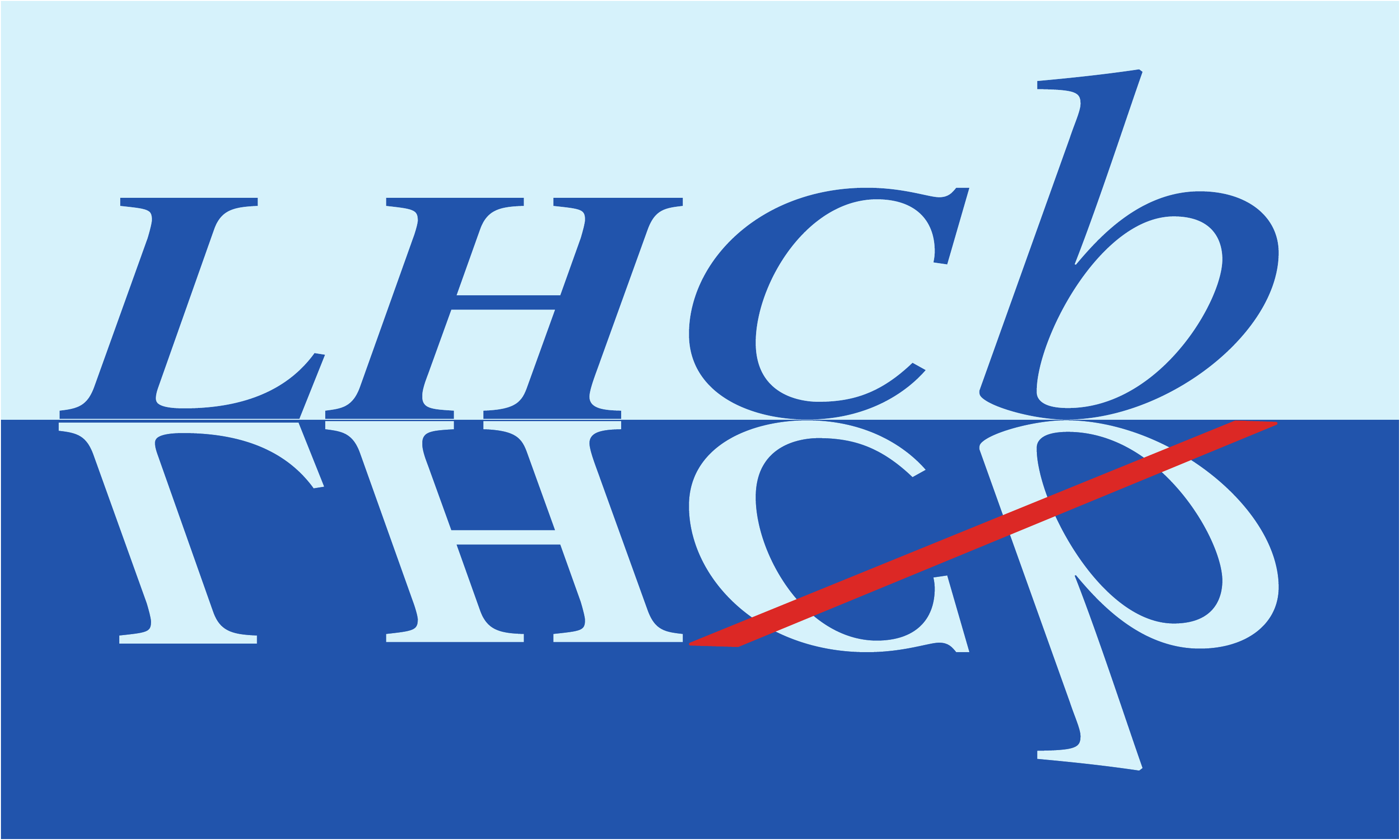}} & &}%
{\vspace*{-1.2cm}\mbox{\!\!\!\includegraphics[width=.12\textwidth]{LHCb/Figures/lhcb-logo.eps}} & &}%
\\
 & & CERN-EP-2026-145 \\
 & & LHCb-PAPER-2025-054 \\
 & & 4 June 2026 \\
 & & \\
\end{tabular*}

\vspace*{3.0cm}

{\normalfont\bfseries\boldmath\huge
\begin{center}
  \papertitle 
\end{center}
}

\vspace*{1.0cm}

\begin{center}      
\paperauthors\footnote{Authors are listed at the end of this paper.}
\end{center}

\vspace{\fill}

\begin{abstract}
This letter reports a study of associated $Z$ boson and prompt \jpsi production in proton-proton collisions at $\sqs =13$\tev with the LHCb detector in the forward rapidity region, using a data sample taken during 2016--2018 corresponding to an integrated luminosity of $5.1\invfb$.
The measured fiducial cross-section is $5.5 \pm 1.5\pb$, significantly exceeding the single-parton scattering expectation of $0.10 \pm 0.08\pb$, indicating that multiparton interactions dominate this process in the explored phase space.
Interpreted within the standard double-parton scattering framework, the data yields an effective cross-section $\sigma_{\rm eff} = 16.6 \pm 4.7\mbarn$.
This provides a direct experimental constraint on the transverse spatial structure of the proton in a kinematic regime characterized simultaneously by small Bjorken-$x$ and an electroweak hard scale set by $Z$ boson mass.
\end{abstract}

\vspace*{2.0cm}

\begin{center}
  Submitted to
  Phys.~Rev.~Lett. 
\end{center}

\vspace{\fill}

{\footnotesize 
\centerline{\copyright~\papercopyright. \href{\paperlicenceurl}{\paperlicence}.}}
\vspace*{2mm}

\end{titlepage}

\newpage
\setcounter{page}{2}
\mbox{~}

\renewcommand{\thefootnote}{\arabic{footnote}}
\setcounter{footnote}{0}

\cleardoublepage

\pagestyle{plain}
\setcounter{page}{1}
\pagenumbering{arabic}

The three-dimensional structure of the proton, encompassing both the longitudinal momentum and transverse spatial distributions of its constituent partons, remains a central topic in quantum chromodynamics (QCD).
While parton distribution functions~(PDF) provide increasingly precise constraints on the longitudinal momentum structure, experimental information on the transverse spatial distribution of partons, including transverse-momentum-dependent distributions~(TMD), remains limited~\cite{Gaunt:2009re,Bartalini:2011jp,Blok:2011bu,Boussarie:2023izj}.
Double-parton scattering (DPS), in which two independent hard partonic interactions occur within a single proton-proton collision, provides sensitivity to the transverse structure of the proton~\cite{Paver:1982yp,Sjostrand:2004pf}.
Within the commonly used factorized approach, the DPS cross-section for the production of final states $A$ and $B$ is expressed as~\cite{Treleani:2007gi}
\begin{equation}
\sigma_{\rm DPS} = \frac{\sigma_A \cdot \sigma_B}{\sigma_{\rm eff}},
\label{equ:eff_xsec}
\end{equation}
where the effective cross-section $\sigma_{\rm eff}$ is related to the transverse overlap function of the colliding protons via $\sigma_{\rm eff}^{-1} = \int \deriv^2b\,F^2(b)$.
Here, $b$ denotes the transverse distance (impact parameter) between the two colliding protons, and $F(b)$ is the transverse overlap function describing the spatial distribution of partons in the proton, normalized such that $\int \deriv^2 b\, F(b) = 1$.
The parameter $\sigma_{\rm eff}$ therefore encodes information on the transverse spatial distribution of partons inside the proton~\cite{Treleani:2007gi,Bartalini:2011jp}.
In the present analysis, the final states $A$ and $B$ correspond to $\decay{\Z}{\mumu}$ and prompt $\decay{\jpsi}{\mumu}$ production, respectively.

Measurements of $\sigma_{\rm eff}$ in a variety of final states have yielded values that are broadly consistent within uncertainties~\cite{Treleani:2007gi}.
At the same time, theoretical studies indicate that $\sigma_{\rm eff}$ may exhibit a non-negligible dependence on the underlying partonic subprocesses, arising from differences in the transverse distributions of valence quarks and gluons~\cite{Blok:2011bu,Egede:2022lws}.
Most existing measurements probe quark-dominated production mechanisms at central rapidities~$y$, motivating complementary studies in gluon-rich kinematic regimes, \ie at small $x$.

The associated production of a $\Z$ boson and a prompt \jpsi meson in proton-proton collisions at $\sqrt{s}=13\tev$ provides a particularly clean laboratory for such studies.
The process benefits from experimentally clean dimuon final states and probes quarkonium production mechanisms in a well-defined electroweak environment~\cite{Bodwin:1994jh,Brambilla:2010cs}.
Within the nonrelativistic QCD framework, production proceeds through both color-singlet and color-octet channels~\cite{Braaten:1999qk,Song:2013uxy,Chao:2012iv}.
In contrast to previous DPS measurements involving quarkonium pairs at relatively low hard scales, such as $\jpsi+\jpsi$, $\jpsi+\psi(2S)$, and $\jpsi+\Upsilonres$ production in the forward region~\cite{LHCB-PAPER-2012-003,LHCB-PAPER-2015-046,LHCb-PAPER-2023-022,LHCb-PAPER-2022-047}, the presence of a $\Z$ boson introduces a characteristic momentum transfer of order $Q^2 \sim m_Z^2$, providing sensitivity to multiparton interactions at a significantly higher hard scale.
The role of the hard scale in DPS processes has been widely discussed in the literature, both experimentally and theoretically, as it may affect the transverse spatial distributions of partons and the degree of factorization between multiple scatterings~\cite{Bartalini:2011jp,Blok:2011bu,Diehl:2011yj}.
Experimental studies involving electroweak bosons in association with heavy-flavor or quarkonium final states at central rapidities have highlighted the importance of accessing higher hard scales in DPS-sensitive observables~\cite{ATLAS:2014ofp,CMS:2026evu,ATLAS:2016ydt,CMS:2014cmt,CMS:2021qsn}.
When combined with the forward rapidity coverage of the \lhcb detector, the $\Z+\jpsi$ final state probes a region of the Bjorken scaling variable $x \sim 10^{-4}$ and hard scale $Q^2 \sim m_Z^2$ that has not been previously constrained experimentally~\cite{Bjorken:1969ja,Bartalini:2011jp}.
In this kinematic regime, the contribution from single parton scattering (SPS) is expected to be suppressed in much of the accessible phase space, 
because observing a high-mass \Z boson and a prompt \jpsi meson in a single partonic interaction requires both dimuon systems to lie within the forward acceptance.
This imposes a strongly asymmetric parton configuration, reducing the available phase space and parton luminosity.
The sensitivity to multiparton dynamics is therefore enhanced in a gluon-dominated environment.

This Letter addresses an experimentally unexplored region of multiparton dynamics by studying the associated production of a $\Z$ boson and a prompt \jpsi meson in the forward region using $pp$ collision data collected with the \lhcb detector during 2016--2018, corresponding to an integrated luminosity of 5.1\invfb.
By combining forward rapidities with an electroweak hard scale, this measurement opens a new window on multiparton interactions in a kinematic regime where existing constraints are scarce.

The \lhcb detector~\cite{LHCb-DP-2008-001,LHCb-DP-2014-002} is a single-arm forward spectrometer covering the pseudorapidity range $2<\eta<5$, designed for the study of particles containing \bquark or \cquark quarks.
Its forward acceptance, together with excellent tracking, vertexing, and muon-identification performance, enables measurements of rare final states involving electroweak bosons and quarkonium in a kinematic region complementary to central detectors.
The online event selection is performed by a trigger system~\cite{LHCb-DP-2012-004}, and the data are processed centrally to deliver analysis-ready samples~\cite{Stripping}.

Simulation is used to model detector acceptance and selection efficiencies.
Proton-proton collisions are generated using \pythia~\cite{Sjostrand:2007gs,Sjostrand:2006za} with an \lhcb-specific configuration~\cite{LHCb-PROC-2010-056}.
Decays of unstable particles are described by \evtgen~\cite{Lange:2001uf} with QED final-state radiation handled by \photos~\cite{davidson2015photos}.
The detector response is simulated using \geant~\cite{Allison:2006ve,Agostinelli:2002hh} as described in Ref.~\cite{LHCb-PROC-2011-006}.
Simulated events are reconstructed and selected with the same algorithms as used for data.

Candidate $\Z+\jpsi$ events are reconstructed in final states containing two pairs of oppositely charged muons, with the \Z boson and \jpsi meson identified through their dimuon decays, \decay{Z}{\mumu} and \decay{\jpsi}{\mumu}, providing a clean experimental signature.
The offline selection criteria follow established strategies used in previous \lhcb analyses~\cite{LHCb-PAPER-2021-037,LHCb-PAPER-2020-011,LHCb-PAPER-2022-047}.
Candidate selection is based on fiducial kinematic requirements, muon and track quality requirements, fit-quality requirements on the individual dimuon vertices, and a combined fit constraining the \Z and \jpsi candidates to a common production vertex; explicit cut definitions are summarized in Appendix~\ref{sec:SM_selection}.
The selection criteria are designed to ensure a clean separation between
genuine DPS configurations and background contributions, in particular 
backgrounds arising from multiple proton-proton interactions occurring in the
same bunch crossing (pileup) 
and from nonprompt quarkonium production.
Compatibility of the reconstructed $\Z+\jpsi$ candidates with a common primary
vertex is quantified using a kinematic fit that constrains the decay 
products to originate from the same vertex~\cite{Hulsbergen:2005pu}.
Candidates with poor fit quality are rejected, efficiently reducing
pileup configurations in which the \Z and \jpsi candidates originate from different
proton-proton collisions, as detailed in Appendix~\ref{sec:app_SM_pileup}.
This requirement also suppresses nonprompt \jpsi mesons from \bquark-hadron
decays, which typically exhibit displaced vertices and poorer compatibility with
the primary vertex~\cite{LHCB-PAPER-2012-003}.

Signal yields are extracted from a simultaneous fit to the mass distributions of the \Z and \jpsi candidates.
Signal shapes are described by analytic functions calibrated using dedicated control samples with similar dimuon topologies, while combinatorial backgrounds are modeled with smooth empirical parameterizations; 
details are given in Appendix~\ref{sec:fit_model}.

\begin{figure*}[tbp]
\centering
\includegraphics[width=0.48\linewidth]{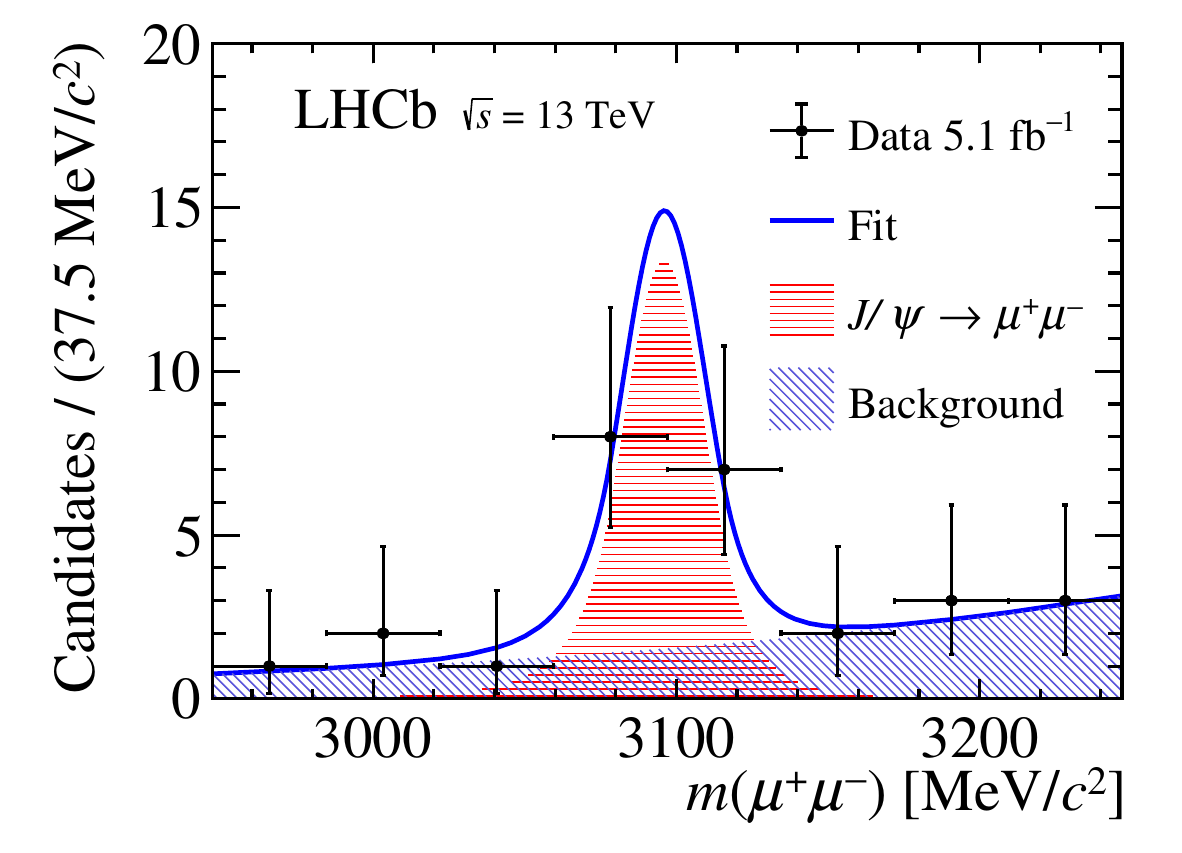}%
\includegraphics[width=0.48\linewidth]{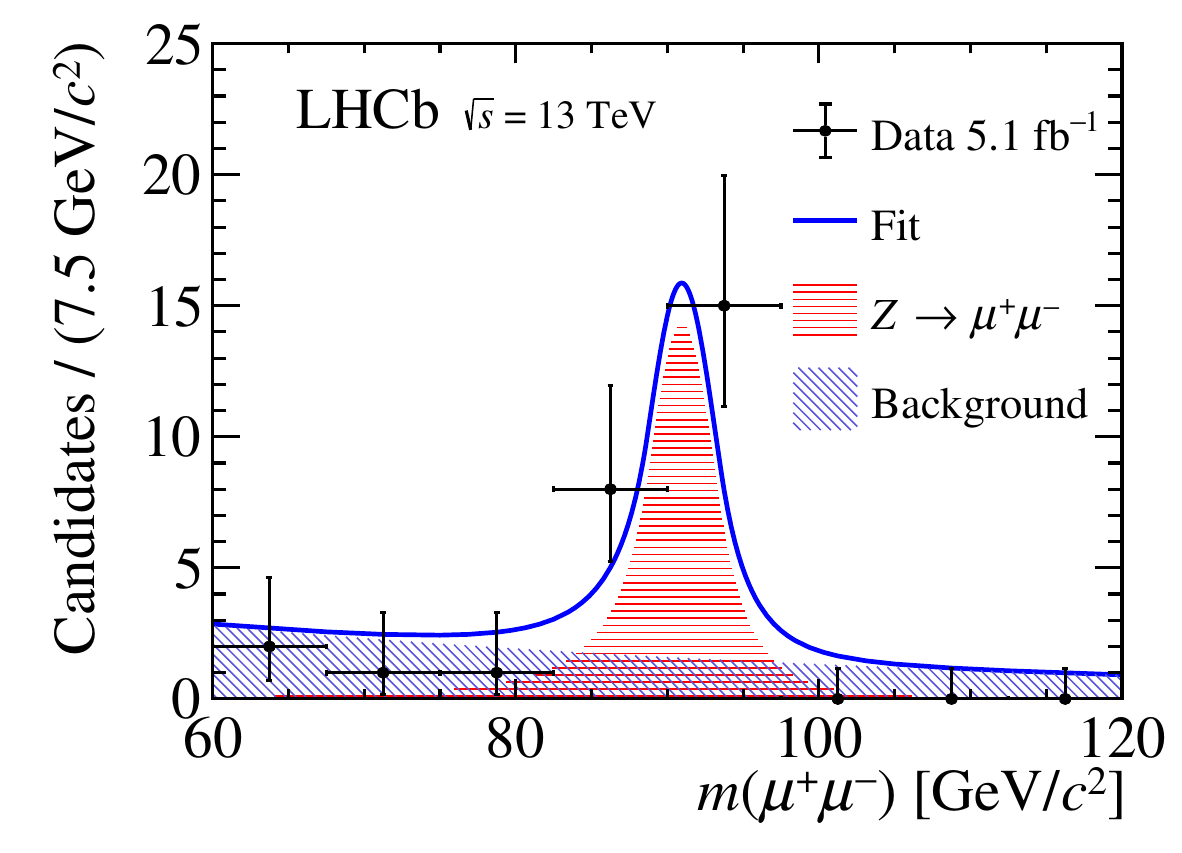}
\caption{Results of a simultaneous fit to the mass distributions in regions centered around the (left)~\jpsi-meson and (right)~\Z-boson masses.}
\label{fig:fitzjpsimass}
\end{figure*}

After background subtraction using the \sPlot technique~\cite{Pivk:2004ty}, per-candidate weights are assigned and propagated through the efficiency correction and cross-section determination.
The results of the simultaneous fit to the mass distributions of the \jpsi and \Z candidates are shown in Fig.~\ref{fig:fitzjpsimass}.

The cross-section for $\Z+\jpsi$ associated production is measured in a fiducial phase-space region defined by the \lhcb forward acceptance for the \decay{Z}{\mumu} and prompt \decay{\jpsi}{\mumu} final states, with $60<m_{\mumu}^{\Z}<120\gevcc$, $\pt^\mu>20\gevc$ and $2<\eta^\mu<4.5$ for the \Z-decay muons, and $0<\jpsipt<14\gevc$ and $2<y^{\jpsi}<4.5$ for the prompt \jpsi candidate. Additional requirements are summarized in Appendix~\ref{sec:SM_selection}. 

It is determined as
\begin{equation}
\sigma^{\Z+\jpsi}
=
\frac{N_{\rm corr}}
{\mathcal{L}\cdot
\BR(\decay{Z}{\mumu})\cdot
\BR(\decay{\jpsi}{\mumu})}\,,
\end{equation}
where $N_{\rm corr}$ is the efficiency-corrected signal yield, $\mathcal{L}$ is the integrated luminosity, and $\BR(\decay{Z}{\mumu})$ and $\BR(\decay{\jpsi}{\mumu})$ are the branching fractions of the decays indicated.

The corrected yield is obtained using per-candidate efficiencies.
The total efficiency is factorized as
\begin{equation}
\epsilon_{\rm tot}
=
\epsilon_{\rm acc}
\cdot
\epsilon_{\rm rec\&sel}
\cdot
\epsilon_{\rm PID}
\cdot
\epsilon_{\rm trig}\,.
\end{equation}
Given the largely uncorrelated reconstruction and selection of the \Z boson and the \jpsi meson, 
the acceptance (acc), reconstruction and selection (rec\&sel), 
particle-identification (PID) and trigger (trig) efficiencies are factorized into products of the 
corresponding single-object efficiencies.
This factorization is essential to ensure a robust interpretation of the
measured cross-section and of the extracted $\sigma_{\rm eff}$ within the DPS
framework.
Efficiencies are determined from simulation and corrected using calibration factors derived from control samples of inclusive \Z and \jpsi production, following standard \lhcb procedures~\cite{LHCb-PAPER-2021-037,LHCB-PAPER-2016-021,LHCB-DP-2013-002,Powell:2010zz}.
The trigger efficiency is defined with respect to the \Z decay products, as the 
candidate selection relies on muons with high transverse momentum, \pt, originating from the \Z decay.
After background subtraction and efficiency correction, the signal yield is $N_{\rm corr}=56\pm16$, where the uncertainty is statistical only.

Systematic uncertainties are evaluated to quantify their impact on the cross-section and on the extracted $\sigma_{\rm eff}$.
The dominant sources arise from signal and background modelling, the stability of the \sPlot-based background subtraction, and efficiency-related corrections.
For the signal and background model component, the signal extraction is repeated with alternative mass-fit parameterisations and control-sample inputs, and the root-mean-square (RMS) of the deviations from the baseline result is assigned as the systematic uncertainty.
The \sPlot-stability uncertainty is evaluated with pseudoexperiments by repeating the fit-and-\sPlot procedure and assigning the RMS of the resulting signal-yield distribution.
Efficiency-related uncertainties are evaluated either by propagation of the efficiency inputs through simulation or, for the PID-binning choice, by taking the larger observed variation from the baseline result.
Independent contributions are added in quadrature.
The integrated luminosity is known to 2.0\%~\cite{LHCB-PAPER-2014-047}.
Branching-fraction uncertainties are taken from Ref.~\cite{PDG2024}.
Alternative signal models are also considered to assess the robustness of the fit.
Uncertainties associated with the inclusive \Z and \jpsi cross-sections used in the determination of $\sigma_{\rm eff}$ follow Refs.~\cite{LHCb-PAPER-2021-037,LHCB-PAPER-2015-037}.
A summary of systematic uncertainties on the $\Z+\jpsi$ cross-section is given in Table~\ref{tab:uncertainty}.

\begin{table}[tpb]
\caption{Summary of relative systematic uncertainties on the $\Z+\jpsi$ cross-section.
The subtotal excluding luminosity corresponds to the systematic uncertainty quoted 
separately from the luminosity uncertainty in the measured cross-section.
For $\sigma_{\rm eff}$, the luminosity uncertainty cancels, while an additional
systematic uncertainty of \(0.9\mbarn\) from the single-\Z and single-\jpsi
cross-section inputs and the SPS subtraction is included.}
\renewcommand{\arraystretch}{1.1}
\label{tab:uncertainty}
\begin{center}
\begin{tabular}{lc}
\hline
\hline
Component & Uncertainty (\%) \\
\hline
Signal and background model & 5.6\\
\sPlot stability & 1.1 \\
MC sample statistics & 0.6 \\
Tracking efficiency & 3.5 \\
PID efficiency & 2.6 \\
$\BR(\decay{\jpsi}{\mumu})$ & 0.6 \\
$\BR(\decay{Z}{\mumu})$ & 0.2 \\
\hline
Subtotal excluding luminosity & 7.2\\
Luminosity & 2.0 \\
\hline
Total including luminosity & 7.5\\
\hline
\hline
\end{tabular}
\end{center}
\end{table}

The SPS contribution to associated $\Z+\jpsi$ production is evaluated using the \textsc{HELAC-Onia}~\cite{Shao:2015vga,Shao:2012iz} package, including both color-singlet and color-octet channels with long-distance matrix elements taken from Ref.~\cite{Lansberg:2019fgm}.
Theoretical uncertainties are estimated by varying the renormalization and factorization scales and by propagating PDF uncertainties~\cite{Dulat:2015mca}.
The DPS contribution is estimated within the standard factorized framework using single-\Z and \jpsi cross-sections as inputs~\cite{Treleani:2007gi}.
The \Z cross-section is obtained with \textsc{MadGraph5\_aMC@NLO}~\cite{Alwall:2014hca, Alwall:2011uj}, while the \jpsi input is taken from \textsc{HELAC-Onia}-based determinations~\cite{Lansberg:2016deg}.
Predictions are quoted for benchmark values of $\sigma_{\rm eff}$ and can be rescaled accordingly.

\begin{figure}[htbp]
\centering
\includegraphics[width=0.5\textwidth]{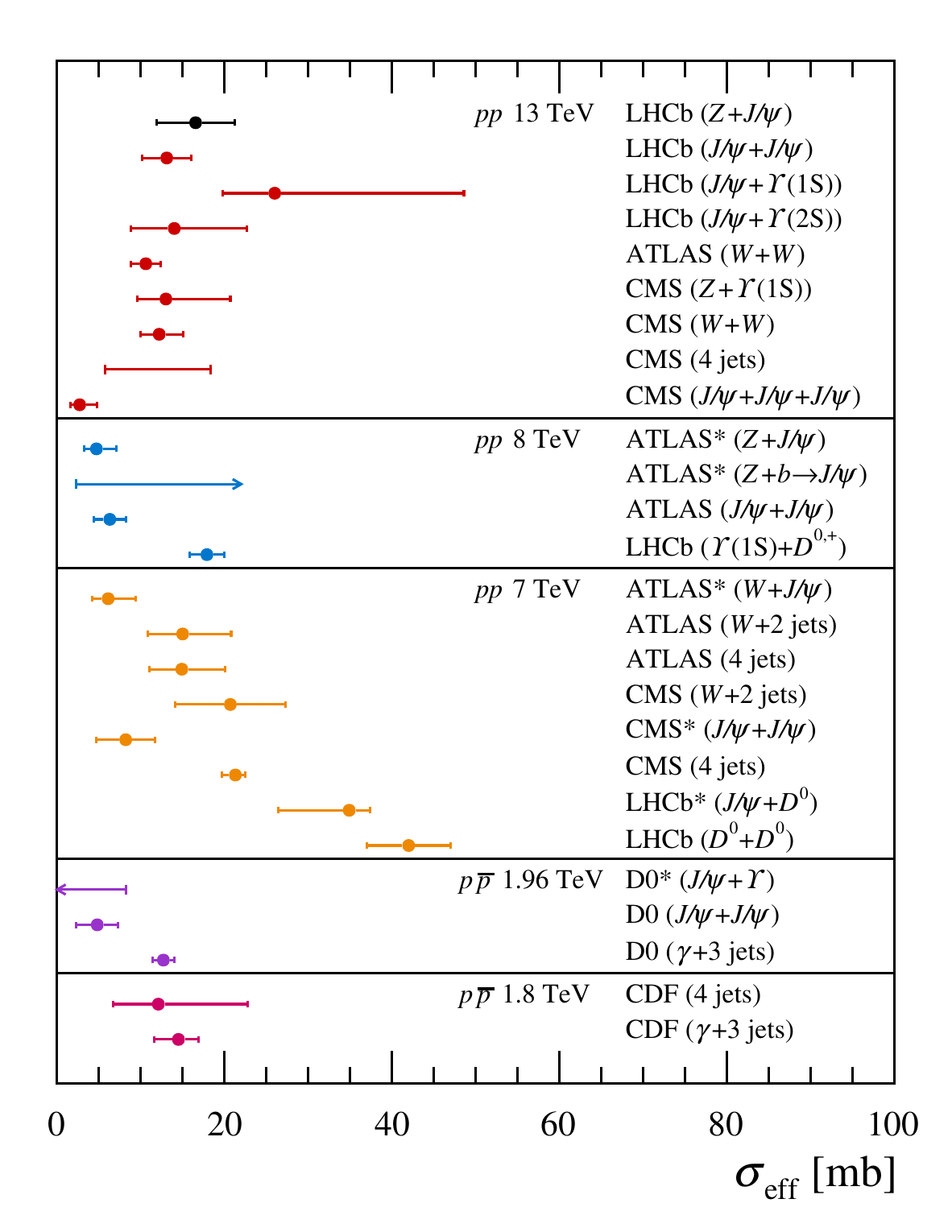}
\caption{Effective cross-sections measured in various particle production processes. The value for $\Z+\jpsi$ from this analysis is shown as a black point. Values marked with an asterisk correspond to theoretical reinterpretations with improved SPS calculations.}
\label{fig:effective_xsec}
\end{figure}

\begin{figure}[htbp]
\centering
\includegraphics[width=0.3\textwidth]{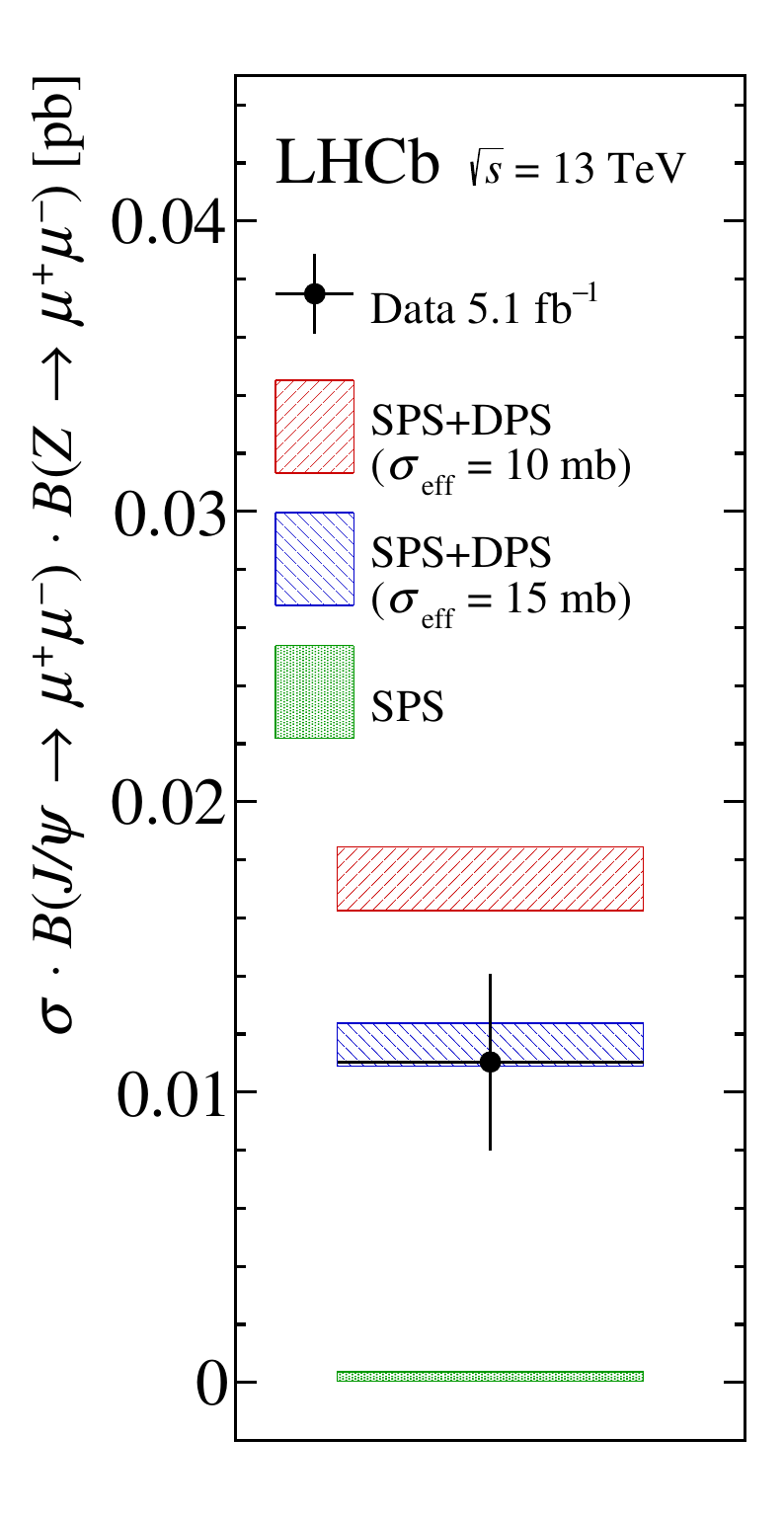}
\caption{Comparison of $\sigma^{\Z+\jpsi}\cdot\BR(\decay{\jpsi}{\mumu})\cdot\BR(\decay{Z}{\mumu})$ between experiment and theoretical predictions. The measurement is shown as a black point, with theoretical results represented by shaded boxes corresponding to SPS predictions from \textsc{HELAC-Onia} and SPS$+$DPS expectations using \textsc{MadGraph5\_aMC@NLO} input for \Z together with \textsc{HELAC-Onia}-based input for \jpsi.}
\label{fig:com_xsec_tot}
\end{figure}

The measured fiducial cross-section for associated $\Z+\jpsi$ production is
\begin{equation*}
\begin{aligned}
\sigma^{\Z+\jpsi}
&=
5.5 \pm 1.5\,(\text{stat}) \pm 0.4\,(\text{syst})\pm 0.1\,(\text{lumi})\pb\,.
\end{aligned}
\end{equation*}
For comparison, the SPS prediction within the same fiducial region is
$\sigma_{\rm SPS}^{\Z+\jpsi} = 0.10 \pm 0.08\pb$,
where the uncertainty reflects renormalization and factorization scale variations and PDF uncertainties.
The measured value therefore significantly exceeds the SPS expectation within the considered theoretical framework, indicating that additional contributions beyond SPS are required to describe $\Z+\jpsi$ production in the forward kinematic region, and providing evidence for a non-SPS contribution in this phase space.

Within the standard DPS factorization approach, the DPS contribution is extracted by subtracting the SPS prediction from the measured cross-section.
The corresponding effective cross-section is then determined using Eq.~(\ref{equ:eff_xsec}) as
\begin{equation*}
\sigma_{\rm eff}
=
16.6\pm4.4\,(\text{stat})
\pm1.5\,(\text{syst})\mbarn\,,
\end{equation*}
where the uncertainty on the luminosity cancels between numerator and denominator.
The systematic uncertainty includes \(1.2\mbarn\) propagated from the nonluminosity 
systematic uncertainty on the measured $\Z+\jpsi$ cross-section and an additional 
\(0.9\mbarn\) from the single-\Z and single-\jpsi cross-section inputs and the SPS subtraction.
The extracted value should be interpreted within the assumptions of the factorized DPS framework and the modelling of the SPS contribution.
In particular, it reflects the effective transverse overlap of partons in a kinematic regime characterized by a hard scale of order $Q^2\sim m_Z^2$ and by forward rapidities corresponding to small-$x$ values.

Figure~\ref{fig:effective_xsec} compares the extracted $\sigma_{\rm eff}$ value with measurements obtained in a broad range of final states and collision energies.
Measurements performed at central rapidities by the ATLAS, CMS, CDF and the D0 experiments predominantly involve electroweak bosons or high-multiplicity-jet final states~\cite{CMS:2021qsn,CMS:2026evu,ATLAS:2025bcb,ATLAS:2014ofp,ATLAS:2016ydt,CMS:2021lxi,ATLAS:2016rnd,CMS:2015wcf,PhysRevD.47.4857,D0:2009apj,PhysRevD.56.3811,ATLAS:2013aph,CMS:2013huw,CMS:2019jcb}.
These processes are characterized by large hard scales of order $Q^2\sim m_W^2$ or $m_Z^2$ and by partonic momentum fractions in the intermediate-$x$ region, with quark-initiated subprocesses providing a substantial contribution.

In contrast, measurements performed at forward rapidity by the \lhcb experiment, such as quarkonium-pair and quarkonium-heavy-flavor production~\cite{LHCb-PAPER-2023-022,LHCb-PAPER-2022-047,LHCB-PAPER-2015-046,LHCB-PAPER-2012-003}, probe significantly lower hard scales and are dominated by gluon-initiated interactions at small $x$.
Despite the very different kinematic regimes and partonic compositions, the corresponding $\sigma_{\rm eff}$ values are found to lie in a similar range, suggesting limited sensitivity of the effective transverse overlap scale to the specific hard process.

The $\Z+\jpsi$ measurement presented here occupies a new region in this context, bringing together the forward rapidity coverage of the \lhcb detector and a hard scale set by the electroweak boson mass.
As illustrated in Fig.~\ref{fig:effective_xsec}, the extracted $\sigma_{\rm eff}$ value is consistent with both the central-rapidity electroweak measurements and the forward quarkonium-based measurements.
This observation provides experimental evidence that the effective transverse scale governing multiparton interactions remains approximately universal across a wide range of $x$ values, hard scales, and partonic initial states, extending existing constraints into the small-$x$, high-$Q^2$ regime.

The product cross-section $\sigma^{\Z+\jpsi}\cdot\BR(\decay{\jpsi}{\mumu})\cdot\BR(\decay{Z}{\mumu})=(1.10\pm0.30)\times10^{-2}\pb$ is compared to SPS and SPS$+$DPS expectations under benchmark assumptions for $\sigma_{\rm eff}$ in Fig.~\ref{fig:com_xsec_tot}.
The SPS contribution alone is found to be strongly suppressed in the forward fiducial region and lies well below the measured value.
In contrast, the DPS component substantially increases the predicted cross-section, bringing the theoretical expectation into agreement with the data.
The comparison illustrates a pronounced sensitivity of the normalization to the assumed value of $\sigma_{\rm eff}$, with values in the range of $\mathcal{O}(10\text{--}20)\,\mathrm{mb}$ providing a reasonable description of the measurement.

\begin{figure*}[tbp]
\centering
\includegraphics[width=0.48\linewidth]{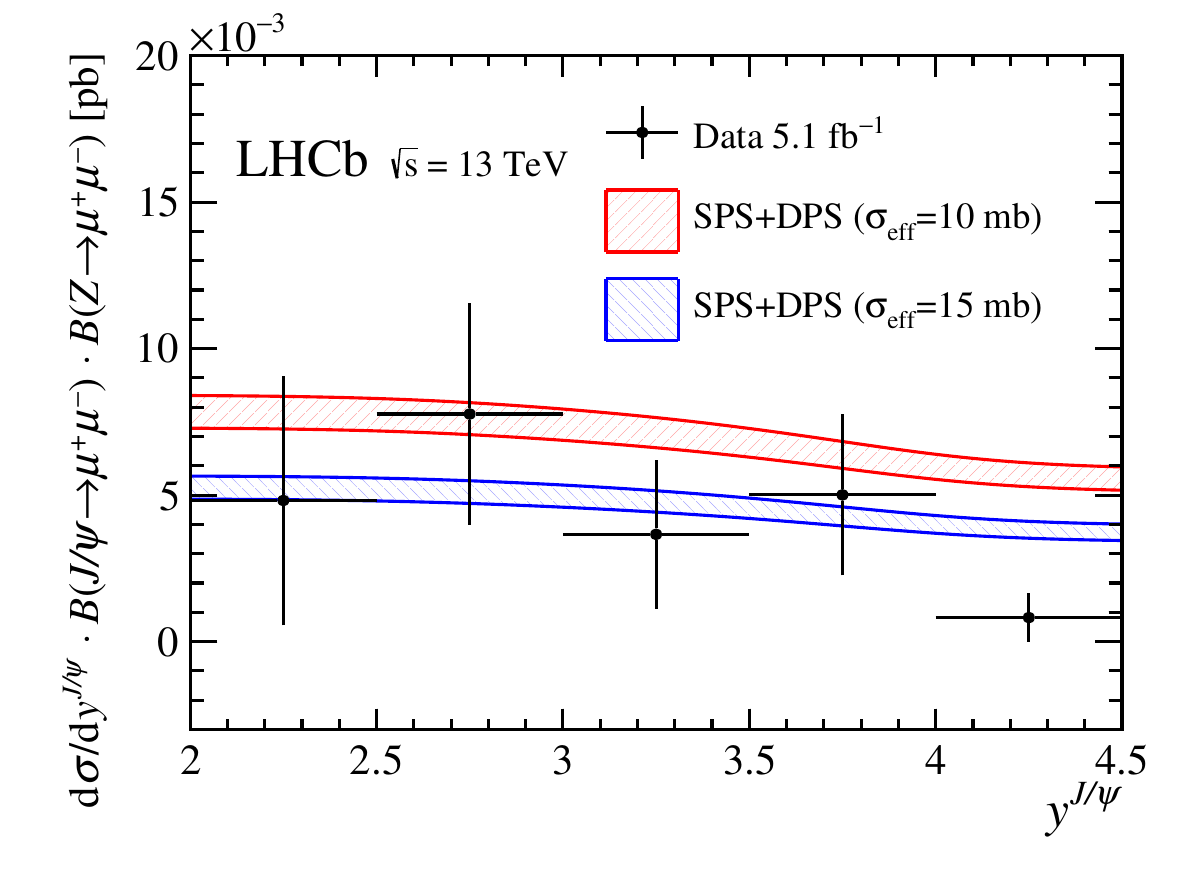}%
\includegraphics[width=0.48\linewidth]{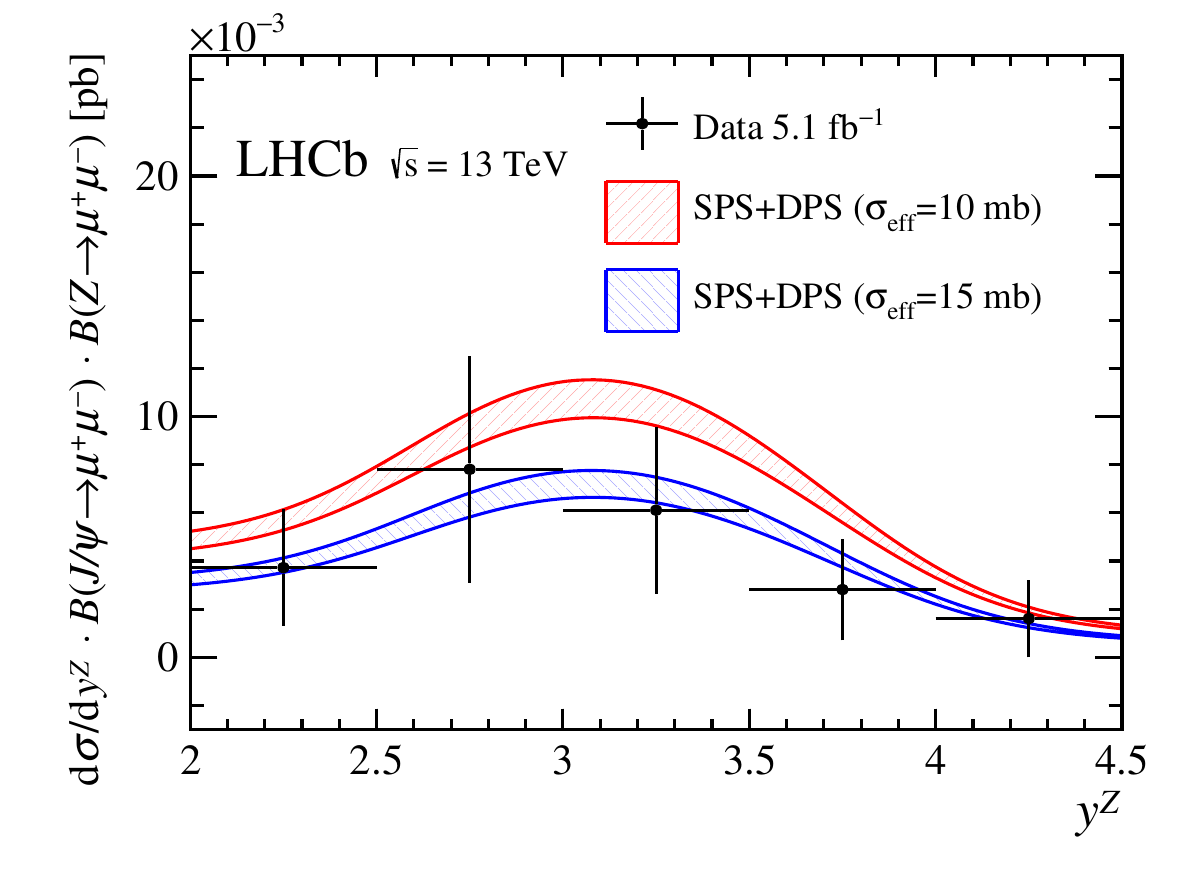}
\caption{Differential cross-sections of $\Z+\jpsi$ production, multiplied by $\BR(\decay{\jpsi}{\mumu})$ and $\BR(\decay{Z}{\mumu})$, compared with SPS$+$DPS predictions
for benchmark values of $\sigma_{\rm eff}=10$ and $15\,\mathrm{mb}$, where the SPS component is evaluated with \textsc{HELAC-Onia}, and the DPS normalization uses single-\Z input from \textsc{MadGraph5\_aMC@NLO} and single-\jpsi input from \textsc{HELAC-Onia}-based determinations.
The distributions are shown as functions of (left) $y^{\jpsi}$ and (right) $y^{\Z}$.
The bands reflect the theoretical uncertainty on the SPS$+$DPS prediction for each
benchmark value of $\sigma_{\rm eff}$.}
\label{fig:com_xsec_rapidity}
\end{figure*}

Differential cross-sections as functions of the rapidities $y^{\Z}$ and $y^{\jpsi}$ are shown in
Fig.~\ref{fig:com_xsec_rapidity} and 
compared with SPS$+$DPS predictions for benchmark values of
$\sigma_{\rm eff}=10$ and $15\,\mathrm{mb}$.
The normalization of the SPS$+$DPS expectation exhibits a pronounced dependence on
$\sigma_{\rm eff}$, while the present data sample does not allow a more
discriminating test of subtle shape differences.
Within the current experimental precision, the measured rapidity distributions are
compatible with the SPS$+$DPS description for values of $\sigma_{\rm eff}$ in the
range favored by the inclusive cross-section.

The $y^{\jpsi}$ and $y^{\Z}$ distributions provide a complementary consistency check of the
DPS interpretation in the forward region.
In particular, the observed distributions show no evidence for strong additional
kinematic correlations beyond those captured by the factorized DPS framework, in a
phase space characterized by forward rapidities and a hard scale set by the electroweak
boson mass.
The corresponding transverse-momentum distributions, which are more sensitive to
soft-QCD modeling at low $\pt$ and are statistically limited in the present dataset,
are provided in Appendix~\ref{sec:app_SM_pt}.

In summary, the associated production of a $\Z$ boson and a prompt \jpsi meson is measured in proton-proton collisions at $\sqs=13\tev$ in the forward region.
The measured cross-section significantly exceeds the SPS expectation in the considered theoretical framework, indicating an essential contribution from multiparton interactions in this phase space.
Interpreted within the standard DPS factorization approach, the data favor a DPS-dominated contribution and allow an effective cross-section of
\mbox{$\sigma_{\rm eff}=16.6\pm4.4\,(\text{stat})\pm1.5\,(\text{syst})\mbarn$}
to be determined.

This result extends experimental constraints on multiparton interactions into a forward, small-$x$, high-$Q^2$ regime, providing a new benchmark for modeling DPS and for constraining the transverse spatial structure of the proton.
More broadly, it helps connect kinematic domains in $x$ and hard scale that had previously remained experimentally separated.

\section*{Acknowledgements}
\noindent 
We thank H.-S. Shao for his invaluable contributions to the discussions on $\Z+\jpsi$ production and for providing the theoretical calculations. 
We express our gratitude to our colleagues in the CERN
accelerator departments for the excellent performance of the LHC. We
thank the technical and administrative staff at the LHCb
institutes.
We acknowledge support from CERN and from the national agencies:
ARC (Australia);
CAPES, CNPq, FAPERJ and FINEP (Brazil); 
MOST and NSFC (China); 
CNRS/IN2P3 and CEA (France);
BMFTR, DFG and MPG (Germany);
INFN (Italy); 
NWO (Netherlands); 
MNiSW and NCN (Poland); 
MEC/IFA (Romania); 
MICIU and AEI (Spain);
SNSF and SER (Switzerland); 
NASU (Ukraine); 
STFC (United Kingdom); 
DOE NP and NSF (USA).
We acknowledge the computing resources that are provided by ARDC (Australia), 
CBPF (Brazil),
CERN, 
IHEP and LZU (China),
IN2P3 (France), 
KIT and DESY (Germany), 
INFN (Italy), 
SURF (Netherlands),
Polish WLCG (Poland),
IFIN-HH (Romania),
PIC (Spain), CSCS (Switzerland), 
GridPP (United Kingdom),
and NSF (USA).
We are indebted to the communities behind the multiple open-source
software packages on which we depend.
Individual groups or members have received support from
RTP (Australia),
FWO Odysseus grant G0ASD25N (Belgium),
Key Research Program of Frontier Sciences of CAS, CAS PIFI, CAS CCEPP (China); 
Minciencias (Colombia);
EPLANET, Marie Sk\l{}odowska-Curie Actions, ERC and NextGenerationEU (European Union);
A*MIDEX, ANR, IPhU and Labex P2IO, and R\'{e}gion Auvergne-Rh\^{o}ne-Alpes (France);
Alexander-von-Humboldt Foundation (Germany);
ICSC (Italy); 
Severo Ochoa and Mar\'ia de Maeztu Units of Excellence, GVA, XuntaGal, GENCAT, InTalent-Inditex and Prog.~Atracci\'on Talento CM (Spain);
the Leverhulme Trust, the Royal Society and UKRI (United Kingdom).

\clearpage

\section*{End matter}
\label{sec:appendix}

\appendix

\section{Event selection summary}
\label{sec:SM_selection}

The \Z candidate must lie within the range $60<m_{\mumu}^{\Z}<120\gevcc$, with its decay products satisfying $\mupt>20\gevc$ and $2<\eta<4.5$, together with muon-identification and momentum-quality requirements. 
The \jpsi candidate is required to be within the ranges $|m_{\mumu}^{\jpsi}-m_{\jpsi}^{\rm PDG}|<500\mevcc$, $0<\jpsipt<14\gevc$ and $2<y^{\jpsi}<4.5$. 
Both \jpsi decay products must satisfy $2.0<\eta<5.0$, $\mupt>650\mevc$ and $\ptot>6\gevc$, along with track-quality and muon-identification requirements that suppress misreconstructed tracks. 
Additional common-vertex fit-quality requirements are applied to the \Z, \jpsi, and combined \Z+\jpsi candidates. 
For the final selected sample, the relevant \Z-side high-$\pt$ single-muon trigger requirements are satisfied with almost 100\% efficiency, thus no additional standalone trigger-efficiency correction is applied.

\section{Pileup-background validation}
\label{sec:app_SM_pileup}

A background contribution arises from configurations in which the reconstructed
$\Z$ and $\jpsi$ candidates originate from different proton-proton interactions
within the same bunch crossing, but are mistakenly associated with a single
reconstructed primary vertex.
Such configurations are referred to as pileup background in the following.
To discriminate this background from genuine $\Z+\jpsi$ production, a
vertex-compatibility goodness-of-fit variable is used.
It is defined as $\log_{10}(\chisqndf)$, where the $\chi^2$ per degree
of freedom is obtained from a fit enforcing a common production vertex for the
reconstructed $\Z$ and $\jpsi$ candidates.
Pileup-like combinations tend to yield poorer vertex compatibility and therefore
larger values of this observable.

\begin{figure}[tbp]

  \centering
  \includegraphics[width=0.48\linewidth]{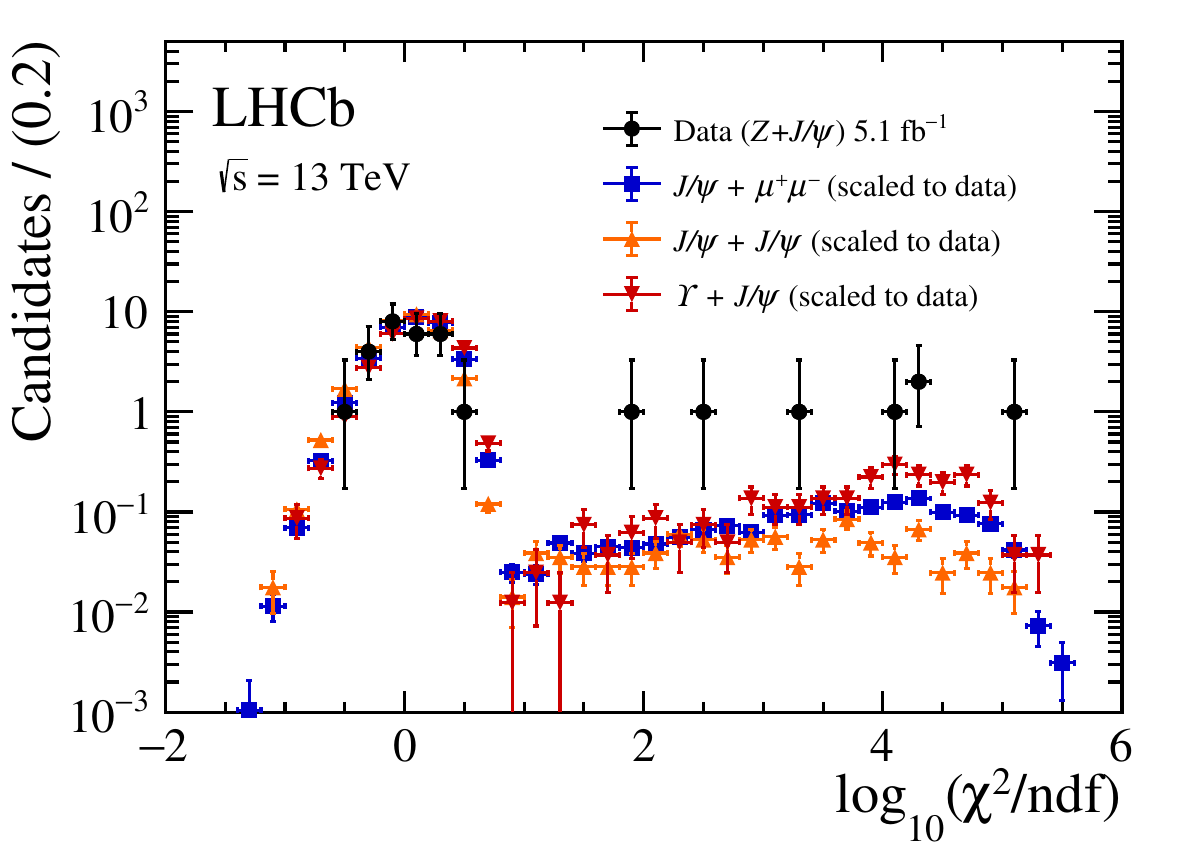}
  \includegraphics[width=0.48\linewidth]{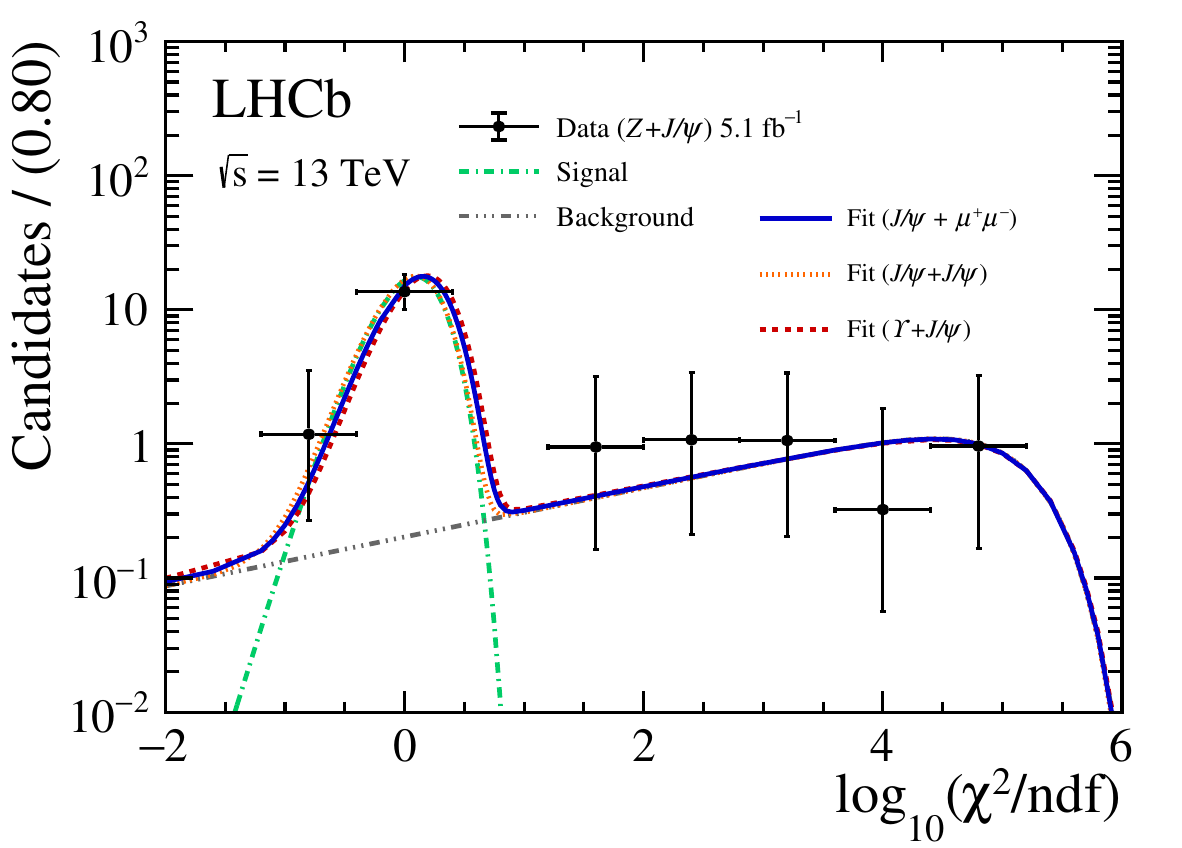}
  \caption{
  (Left) Normalized $\log_{10}(\chisqndf)$ distributions for $\Z+\jpsi$
  data and the $\jpsi+\mu^+\mu^-$, $\jpsi+\jpsi$, and $\jpsi+\Upsilonres$ large control samples. The control samples are scaled
  to the same integral as the data to facilitate direct shape comparison.
  (Right) Fits to the $\Z+\jpsi$ $\log_{10}(\chisqndf)$ distribution
  obtained by fixing the signal-shape parameters to values determined from
  different control samples, yielding $\chisqndf=0.96$. The parameterization derived from the
  $\jpsi+\jpsi$ control sample is shown together with alternative choices. The
  signal and background components of this fit are displayed separately.
  }
  \label{fig:app_SM_dtfchi2_compare}
\end{figure}

Given the limited size of the $\Z+\jpsi$ signal sample, the shape of the 
$\log_{10}(\chisqndf)$ distribution cannot be reliably determined from 
the signal region alone.
Instead, large control samples are used to validate the stability
of the distribution shape across different event topologies.
Figure~\ref{fig:app_SM_dtfchi2_compare} compares the normalized
$\log_{10}(\chisqndf)$ distributions observed in the $\jpsi+\mu^+\mu^-$, $\jpsi+\jpsi$, and $\jpsi+\Upsilonres$ control samples
and in the $\Z+\jpsi$ data.
The control-sample histograms are scaled to the same total yield as the
$\Z+\jpsi$ sample to allow direct shape comparison.
The consistent peak structure and similar overall behavior across samples support
the use of control-sample-derived shape information in the $\Z+\jpsi$ analysis.

To quantify the impact of the control-sample choice, the $\Z+\jpsi$
$\log_{10}(\chisqndf)$ distribution is fitted multiple times,
each time fixing the signal shape parameters to those obtained from
a different control sample.
In these fits, only the relative normalization of the prompt-like and
pileup-related components is free to vary.
The resulting fits, shown in Fig.~\ref{fig:app_SM_dtfchi2_compare} (right),
provide comparable descriptions of the data.
The variations observed across control-sample parameterizations are small
compared to the statistical uncertainty of the $\Z+\jpsi$ sample and are
propagated as a systematic uncertainty associated with the pileup-shape choice in
the main analysis.

\section{Fit model details}
\label{sec:fit_model}

The signal components in the mass shapes are modeled by double-sided Crystal Ball functions~\cite{Skwarnicki:1986xj}.
Because of the limited size of the $\Z+\jpsi$ signal sample, the shape parameters
used in the baseline fit are calibrated with large control samples
of inclusive \jpsi and \Z production.
The combinatorial background contribution is parameterized with smooth exponential functions.

The raw signal yield obtained from the baseline fit is
$14.4 \pm 3.8$ candidates.
The efficiency-corrected yield is computed event-by-event as a weighted sum over
selected candidates,
\[
N_{\rm corr} = \sum_i \frac{\omega_i}{\epsilon_i},
\]
where $\omega_i$ denotes the per-candidate signal weight after background
subtraction and $\epsilon_i$ is the corresponding total efficiency evaluated from
the efficiency maps.
Its statistical uncertainty is propagated from the same event-by-event weighted
sum,
\[
\sigma(N_{\rm corr}) =
\sqrt{\sum_i \left(\frac{\omega_i}{\epsilon_i}\right)^2 }.
\]
For the present dataset, this gives $N_{\rm corr}=56\pm16$.

The systematic uncertainty associated with the signal and background model is evaluated by repeating the signal extraction with alternative mass-fit parameterizations and control-sample inputs for the \jpsi and \Z mass shapes. 
The RMS of the deviations of the extracted signal yield from the baseline result is assigned as the model uncertainty. The pileup-shape contribution is evaluated analogously using alternative control-sample parameterizations of the common-vertex fit-quality distribution, and is combined in quadrature with the combinatorial-background modeling contribution. The systematic uncertainty from the mass fit is evaluated with pseudoexperiments generated from the baseline mass model: each pseudoexperiment is fitted, the \sPlot procedure is applied, and the RMS of the resulting signal-yield distribution is assigned as the uncertainty.

\section{Auxiliary transverse-momentum distributions}
\label{sec:app_SM_pt}

\begin{figure}[tbp]
\centering
\includegraphics[width=0.48\linewidth]{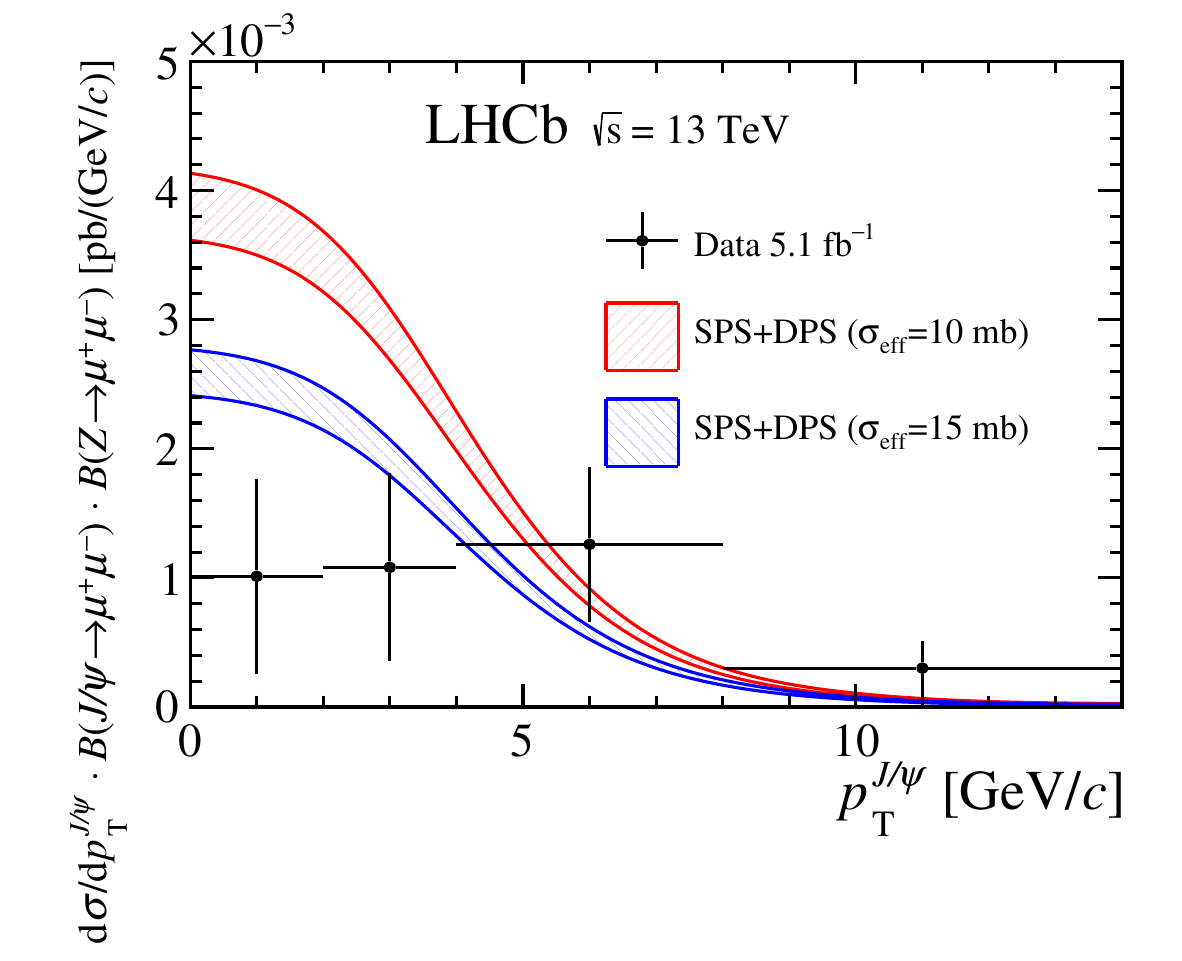}%
\includegraphics[width=0.48\linewidth]{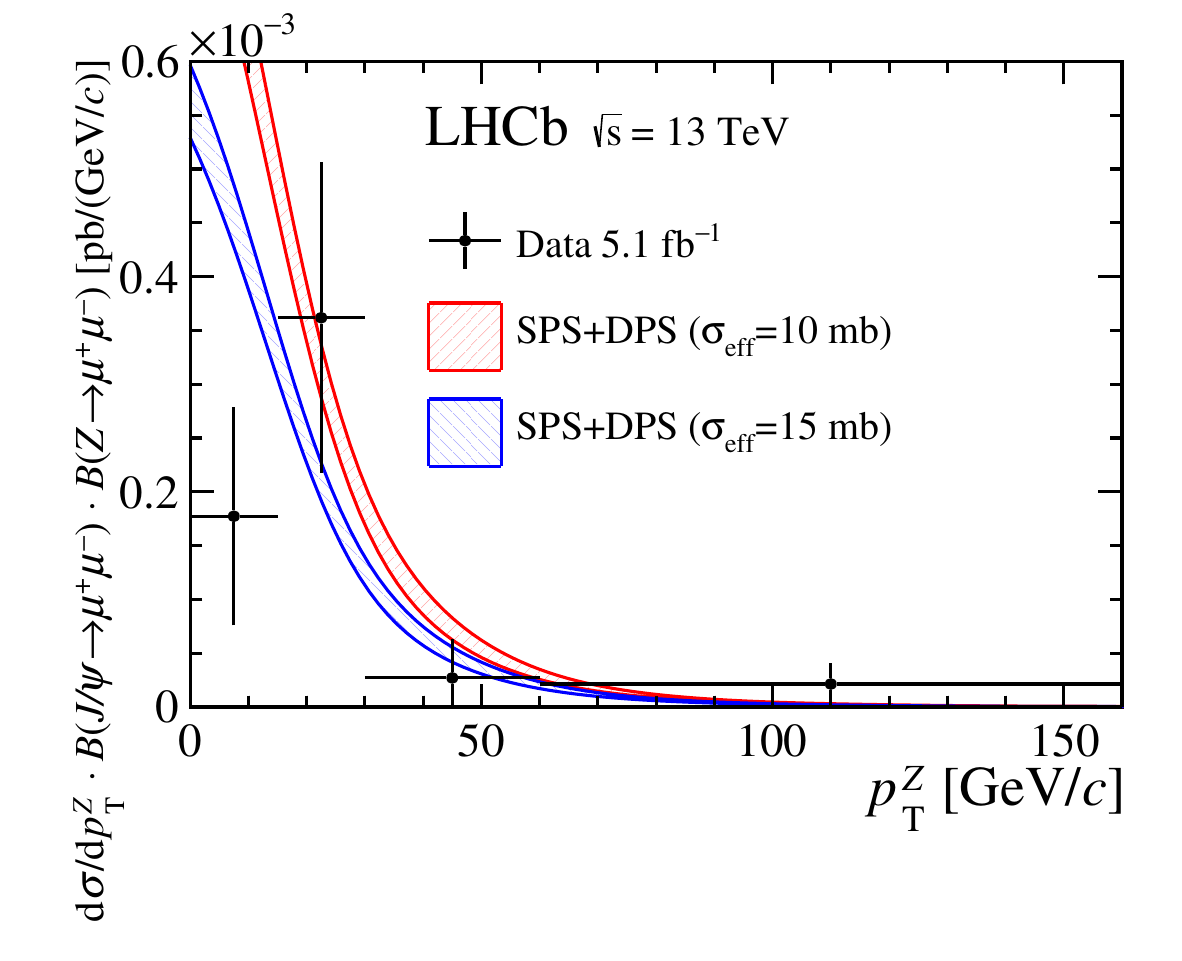}
\caption{Differential cross-sections of $\Z+\jpsi$ production, multiplied by
$\BR(\decay{\jpsi}{\mumu})$ and $\BR(\decay{\Z}{\mumu})$, compared with SPS$+$DPS predictions
for benchmark values of $\sigma_{\rm eff}=10\,\mathrm{mb}$ and $\sigma_{\rm eff}=15\,\mathrm{mb}$, shown as
functions of (left) $\jpsipt$ and (right) $\zpt$.}
  \label{fig:app_com_xsec_pt_supp}
\end{figure}

Additional differential cross-sections in transverse-momentum observables are
provided in Fig.~\ref{fig:app_com_xsec_pt_supp}.
These distributions are not included in the main text, where the emphasis is
placed on the observables most directly connected to the forward-rapidity
kinematics.
They provide an auxiliary consistency check of the SPS$+$DPS description for
benchmark values of $\sigma_{\rm eff}=10\,\mathrm{mb}$ and $\sigma_{\rm eff}=15\,\mathrm{mb}$ in a
statistics-limited regime.
Within the current experimental uncertainties, the data are compatible with the
benchmark predictions and do not provide additional constraining power on
$\sigma_{\rm eff}$ beyond what is reported in the main text.

\clearpage

\clearpage

\addcontentsline{toc}{section}{References}
\ifx\mcitethebibliography\mciteundefinedmacro
\PackageError{LHCb.bst}{mciteplus.sty has not been loaded}
{This bibstyle requires the use of the mciteplus package.}\fi
\providecommand{\href}[2]{#2}

\clearpage
\centerline
{\large\bf LHCb collaboration}
\begin
{flushleft}
\small
R.~Aaij$^{38}$\lhcborcid{0000-0003-0533-1952},
A.S.W.~Abdelmotteleb$^{58}$\lhcborcid{0000-0001-7905-0542},
C.~Abellan~Beteta$^{52}$\lhcborcid{0009-0009-0869-6798},
F.~Abudin\'en$^{58}$\lhcborcid{0000-0002-6737-3528},
T.~Ackernley$^{62}$\lhcborcid{0000-0002-5951-3498},
A. A. ~Adefisoye$^{70}$\lhcborcid{0000-0003-2448-1550},
B.~Adeva$^{48}$\lhcborcid{0000-0001-9756-3712},
M.~Adinolfi$^{56}$\lhcborcid{0000-0002-1326-1264},
P.~Adlarson$^{86}$\lhcborcid{0000-0001-6280-3851},
C.~Agapopoulou$^{14}$\lhcborcid{0000-0002-2368-0147}
% ...

{\footnotesize \it

% ...
$^{14}$Universit{\'e} Paris-Saclay, CNRS/IN2P3, IJCLab, Orsay, France\\
% ...
$^{38}$Nikhef National Institute for Subatomic Physics, Amsterdam, Netherlands\\
% ...
$^{48}$Instituto Galego de F{\'\i}sica de Altas Enerx{\'\i}as (IGFAE), Universidade de Santiago de Compostela, Santiago de Compostela, Spain\\
% ...
$^{52}$Physik-Institut, Universit{\"a}t Z{\"u}rich, Z{\"u}rich, Switzerland\\
% ...
$^{56}$H.H. Wills Physics Laboratory, University of Bristol, Bristol, United Kingdom\\
% ...
$^{58}$Department of Physics, University of Warwick, Coventry, United Kingdom\\
% ...
$^{61}$School of Physics and Astronomy, University of Glasgow, Glasgow, United Kingdom\\
$^{62}$Oliver Lodge Laboratory, University of Liverpool, Liverpool, United Kingdom\\
% ...
$^{70}$Syracuse University, Syracuse, NY, United States\\
% ...
$^{86}$Department of Physics and Astronomy, Uppsala University, Uppsala, Sweden, associated to $^{61}$\\
% ...
}
\end{flushleft}\end{document}